\documentclass[12pt]{article}
\usepackage{amsfonts}
\usepackage{amssymb}
\usepackage{amscd}

\begin{document}

\title{\large{\bf ON GENERALIZED FUNCTIONS IN ADELIC QUANTUM MECHANICS}}

\bigskip

\author{ {\bf Branko Dragovich} \\ Institute of Physics, P. O. Box 57 \\ 11001 Belgrade, Yugoslavia}

\date{}
\bigskip

\bigskip

\bigskip

\maketitle

\begin{abstract}
{Some aspects of adelic generalized functions, as linear
continuous functionals on the space of Schwartz-Bruhat functions,
are considered. The importance of adelic generalized functions in
adelic quantum mechanics is demonstrated. In particular, adelic
product formula for Gauss integrals is derived, and the connection
between the functional relation for the Riemann zeta function and
quantum states of the harmonic oscillator is stated.}
\end{abstract}

\bigskip



\bigskip
\bigskip
\noindent{\bf 1.\ \ Introduction}
\bigskip

Generalizations in mathematics and physics are usually very
fruitful tools which extend a topic and also lead to new insights
into the subject itself. According to the title of this article we
shall consider generalized functions of adelic arguments. Since,
in a sense, adeles are a generalization of rational numbers it is
useful to start with the field of rational numbers ${\mathbb Q}$.

The Ostrowski theorem says that each norm (valuation) on ${\mathbb
Q}$ is equivalent to the absolute value $|\, \,|_\infty$ or to
some $p$-adic norm $|\, \,|_p$\,. On the basis of definition,
$|p^\nu \frac{s}{t}|_p = p^{-\nu}  \, (0\neq t, s, \nu \in
{\mathbb Z} ; s, t$ are not divisible by prime number $p$ ) and
$|0|_p =0$, $p$-adic norm is the non-archimedean one, {\it i.e.}
$|x+y|_p \leq \mbox{max} (|x|_p\, , |y|_p)$. For every $p$ there
is one $p$-adic norm on ${\mathbb Q}$. As the field of real
numbers ${\mathbb R}\equiv {\mathbb Q}_\infty$, so the field of
$p$-adic numbers ${\mathbb Q}_p$ \cite{schikhof} is an analogous
completion of ${\mathbb Q}$ with respect to the norm $|\, \,|_p$ .

Any $x \in {\mathbb Q}_p$ can be presented in the form
$$
x = p^\nu \, (x_0 + x_1 \, p + x_2\, p^2 + \cdots) , \quad \nu \in
{\mathbb Z} , \quad x_i = 0, 1, \cdots , p-1 \, .  \eqno(1)
$$
It is of special interest the ring of $p$-adic integers
 ${\mathbb Z}_p = \{x \in {\mathbb Q}_p :\, |x|_p \leq 1  \}$, {\it i.e.}
such numbers have exponent $\nu \geq 0$ in the canonical expansion
(1).

Let ${\mathcal P}$ be any set of finite prime numbers and let
$$
{\mathbb A} ({\mathcal P}) = {\mathbb Q}_\infty \times \prod_{p
\in {\mathcal P}} {\mathbb Q}_p \times \prod_{p \notin {\mathcal
P}} {\mathbb Z}_p \, , \quad {\mathbb A} =\bigcup_{\mathcal P}
{\mathbb A} ({\mathcal P}) \, . \eqno(2)
$$
If $ {\mathcal P}\subset {\mathcal P}' $ it holds  $ {\mathbb A}
({\mathcal P}) \subset {\mathbb A} ({\mathcal P}')$ and $
\bigcup_{\mathcal P} {\mathbb A} ({\mathcal P}) =
\bigcup_{{\mathcal P}'} {\mathbb A} ({\mathcal P}')$. $ {\mathbb
A}$ is the space of adeles \cite{platonov}, \cite{gelfand}. Any
adele $a \in {\mathbb A} $ is an infinite sequence
$$
 a = (a_\infty , \, a_2 , \cdots , a_p\, , \cdots)     \eqno(3)
$$
such that $|a_p|_p \leq 1$ for all but a finite number of $p$,
{\it i.e.} it may be $|a_p|_p > 1$ only when $p \in {\mathcal P}$.
${\mathbb A}$ is a topological ring.

The subset of ${\mathbb A}$ with its multiplicative inverses is
called the group of ideles ${\mathbb A}^\ast$. In fact,
$$
{\mathbb A}^\ast = \bigcup_{\mathcal P} {\mathbb A}^\ast
({\mathcal P}) \, , \quad {\mathbb A}^\ast ({\mathcal P}) =
{\mathbb Q}_\infty^\ast \times \prod_{p \in {\mathcal P}} {\mathbb
Q}_p^\ast \times \prod_{p \notin {\mathcal P}} {U}_p \, ,
\eqno(4)
$$
where ${\mathbb Q}_\infty^\ast = {\mathbb Q}_\infty \setminus \{
0\} \, , {\mathbb Q}_p^\ast = {\mathbb Q}_p \setminus \{ 0\}$ and
$U_p = \{ x \in {\mathbb Q}_p  \, : |x|_p = 1\}$ is the set of
$p$-adic units. Any idele $\lambda \in {\mathbb A}^\ast$ can be
presented in the form
$$
 \lambda = (\lambda_\infty , \, \lambda_2 , \cdots , \lambda_p \, , \cdots)
$$
with restriction $\lambda_\infty \neq 0\, , \lambda_p \neq 0$, and
$| \lambda_p| \neq 1$ only for $p\in {\mathcal P}$. Adelic
sequence $ (r , \, r , \cdots , r \cdots)$ is a principal adele if
$r \in {\mathbb Q}$, and it is a principal idele if  $r \in
{\mathbb Q}^\ast$. ${\mathbb A}$ and ${\mathbb A}^\ast$ have the
corresponding invariant Haar measures:
$$
dx = dx_\infty \, dx_2\, \cdots \, dx_p \, \cdots  \quad
\mbox{and} \quad  d^\ast x = d^\ast x_\infty \, d^\ast x_2\,
\cdots \, d^\ast x_p \, \cdots \, ,
$$
respectively. It holds: $d^\ast x_\infty =
|x_\infty|_\infty^{-1}\, dx_\infty$ and  $d^\ast x_p = (1-
p^{-1})^{-1} \, |x_p|_p^{-1} \, dx_p$ for each $p$.

There are two kinds of analysis over ${\mathbb A}$ induced by the
map $(i)$: $F : {\mathbb A} \longrightarrow {\mathbb A}$ and the
map $(ii)$ $f : {\mathbb A} \longrightarrow {\mathbb C}$, where
${\mathbb C}$ is the standard field of complex numbers. An
analogous statement has a place for ${\mathbb A}^\ast$. Let us
mention some functions of type $(ii)$. An additive character on
${\mathbb A}$ is
$$
\chi (x) = \chi_\infty (x_\infty)\, \prod_p \chi_p (x_p) = \exp{(-
2 \pi i x_\infty)} \, \prod_p \exp{( 2 \pi i \{x_p \}_p)}\, ,
\eqno(5)
$$
where $\{x_p \}_p$ means the fractional part of $x_p$ in the
canonical expansion (1).

Function
$$
\pi_\alpha (\lambda) = |\lambda_\infty|_\infty^\alpha \, \prod_p
|\lambda_p|_p^\alpha \,\, , \quad \lambda \in {\mathbb A}^\ast \,
\, , \quad \alpha \in {\mathbb C}        \eqno(6)
$$
is a multiplicative character on ${\mathbb A}^\ast$. One can show
that $\chi (x) = 1 \quad (\pi (x) = 1)$ if $x$ is a principal
adele (idele).

Maps $\varphi_{\mathcal P} : {\mathbb A}\longrightarrow {\mathbb
C}$ which have the form
$$
\varphi_{\mathcal P} (x) = \varphi_{\infty} (x_\infty) \, \prod_{p
\in {\mathcal P}} \varphi_{p} (x_p) \, \prod_{p\notin {\mathcal
P}} \Omega_p (|x_p|_p) \, ,              \eqno(7)
$$
where $\varphi_{\infty} (x_\infty) \in {\mathcal S} ({\mathbb
Q_\infty})$ are infinitely differentiable functions on ${\mathbb
Q}_\infty$ and fall to zero faster than any power of
$|x_\infty|_\infty$ as $|x_\infty|_\infty \rightarrow \infty \, ,
\quad  \varphi _p (x_p) \in {\mathcal D}({\mathbb Q}_p)$ are
locally constant functions with compact support, and
$$
\Omega_p \, (|x_p|_p) = \left\{  \begin{array}{ll}
                 1,   &   \,\,\,\, |x_p|_p  \leq 1 \, ,  \\
                 0,   &   \,\,\,\,  |x_p|_p > 1 \, ,
                 \end{array}    \right.   \eqno(8)
$$
are called elementary functions on ${\mathbb A}$. Finite linear
combinations of elementary functions (7) make the set of
Schwartz-Bruhat functions ${\mathcal S}({\mathbb A})$. The Fourier
transform is
$$
\tilde{\varphi} (\xi) = \int_{\mathbb A} \varphi (x) \, \chi (\xi
x)\, dx                                    \eqno(9)
$$
and it maps one-to-one ${\mathcal S}({\mathbb A})$ onto ${\mathcal
S}({\mathbb A})$.

This paper is devoted to adelic generalized functions as linear
continuous functionals on ${\mathcal S}({\mathbb A})$, which plays
the role of space of test functions. Some aspects of these
functions will be considered in the context of adelic quantum
mechanics \cite{dragovich1}.

\newpage

\bigskip
\noindent{\bf 2.\ \ Adelic  generalized functions}
\bigskip

Let us introduce adelic generalized functions as an extension of
this concept in the real \cite{vladimirov1} and $p$-adic
\cite{vladimirov2} (see also \cite{gelfand}, \cite{bruhat} and
\cite{volovich1}) case.

\bigskip
\noindent{\bf Definition.} Adelic generalized functions are linear
continuous functionals on the space of Schwartz-Bruhat functions $
{\mathcal S}({\mathbb A})$.

\bigskip

We shall denote the set of adelic generalized functions by
${\mathcal S}'({\mathbb A})$. The corresponding functionals $(f ,
\varphi) \in {\mathbb C}$, where $\varphi \in {\mathcal
S}({\mathbb A})$ and $f \in {\mathcal S}'({\mathbb A})$, may be
presented in the form
$$
(f , \varphi) = \sum_{i=1}^n C_i \, (f , \varphi_{{\mathcal
P}_i}^{(i)} ) \, ,         \eqno(10)
$$
$$
(f , \varphi_{{\mathcal P}_i}^{(i)}) = (f_\infty ,
\varphi_{\infty}^{(i)}) \, \prod_{p\in {\mathcal P}_i} (f_p ,
\varphi_{p}^{(i)}) \, \prod_{p\notin {\mathcal P}_i} (f_p ,
\Omega_p) \, .           \eqno(11)
$$
Formula (10) follows from the fact that $\varphi \in {\mathcal
S}({\mathbb A}) $ may be presented as $\varphi (x) =\sum_{i=1}^n
C_i \, \varphi_{{\mathcal P}_i}^{(i)}  (x)$, where
$\varphi_{{\mathcal P}_i}^{(i)}  (x)$ has the form (7). Of course,
one demands convergence of the infinite product in (11). Thus, an
adelic generalized function $f \in {\mathcal S}'({\mathbb A}) $ is
composed of terms with infinite products (11) of generalized
functions $f_\infty \in {\mathcal S}'({\mathbb Q}_\infty)$ and
those $f_p \in {\mathcal D}' ({\mathbb Q}_p)$ for which $\prod_p
(f_p , \Omega_p)$ is convergent. Adelic generalized function $f$
is a continuous functional on ${\mathcal S}({\mathbb A})$ in the
sense that $(f , \varphi_k ) \rightarrow (f , \varphi)$ if $
\varphi_k \rightarrow \varphi \in {\mathcal S}({\mathbb A})$ when
$k \rightarrow \infty$.

Let us write down some examples of adelic generalized functions.

\bigskip

\noindent{\bf Example 1.} ${\mathcal S}({\mathbb A}) \subset
{\mathcal S}'({\mathbb A})$.

Namely for $f\, , \varphi \in {\mathcal S}({\mathbb A})$ we have
$$
\int_{{\mathbb A}} f (x)\, \varphi (x) \, dx =\sum_{i=1}^n \,
\sum_{j=1}^m C_i \, C_j \, \int_{\mathbb A}  \varphi_{{\mathcal
P}_i}^{(i)} (x) \, \varphi_{{\mathcal P}_j}^{(j)} (x) \, dx
\eqno(12)
$$
and it is convergent. It is worth noting that integration over
${\mathbb A}$ in (12) is practically reduced to integrals over
${\mathbb A} ({\mathcal P}_{i j})$, where ${\mathcal P}_{i j} =
{\mathcal P}_{i }\bigcup {\mathcal P}_{ j}$ .

\bigskip
\noindent{\bf Example 2.} Additive character $\chi \in {\mathcal
S}'({\mathbb A})$.

It is enough to see that the Fourier transform
$$
\int_{\mathbb A} \varphi_{\mathcal P} (x) \, \chi (x) \, dx =
\tilde{\varphi}_\infty (1) \, \prod_{p \in {\mathcal P}}
\tilde{\varphi}_p (1) \, \prod_{p \notin {\mathcal P}} \Omega_p
(|1|_p) = \tilde{\varphi}_\infty (1) \, \prod_{p \in {\mathcal P}}
\tilde{\varphi}_p (1)              \eqno(13)
$$
is a finite product.

\bigskip

\noindent{\bf Example 3.} With respect to the Haar measure $d^\ast
x$ the multiplicative character $\pi_\alpha \in {\mathcal
S}'({\mathbb A})$.

Indeed, one has
$$
\Phi_{\mathcal P} (\alpha) = \int_{{\mathbb A}^\ast} |x|^\alpha \,
\varphi_{\mathcal P} (x)\, d^\ast x = \int_{{\mathbb Q}_\infty}
|x_\infty|_\infty^{\alpha -1} \, \varphi_\infty (x_\infty)\,
dx_\infty
$$
$$
\times \prod_{p\in {\mathcal P}} \frac{1-p^{-\alpha}}{1-p^{-1}}\,
\int_{{\mathbb Q}_p} |x_p|_p^{\alpha -1} \, \varphi_p (x_p) \,
dx_p \, \, \, \zeta (\alpha) \, , \quad \mbox{Re}\,\, \alpha > 1\,
, \eqno(14)
$$
where $\zeta (\alpha)$ is the Riemann zeta function.

\bigskip

\noindent{\bf Example 4.} Locally integrable function $f \in
L^1_{loc} ({\mathbb A})$ induces an adelic generalized function by
formula
$$
 (f , \varphi) = \int_{\mathbb A} f (x) \, \varphi (x) \, dx \, ,
 \quad \varphi \in {\mathcal S}({\mathbb A}) \, .
$$

\bigskip

\noindent{\bf Example 5.}  The Dirac $\delta$-function on
${\mathbb A}$, where
$$
\delta (x) = \delta_\infty (x_\infty)\, \prod_p \delta_p (x_p) =
\int_{{\mathbb Q}_\infty} \chi_\infty (x_\infty y_\infty) \,
dy_\infty \, \, \prod_p \int_{{\mathbb Q}_p} \chi_p (x_p\, y_p) \,
dy_p \, , \eqno(15)
$$
is an adelic generalized function, {\it i.e.} $\, \delta\in
{\mathcal S}'({\mathbb A}) $.

Namely,
$$
(\delta , \varphi_{\mathcal P}) = (\delta_\infty , \varphi_\infty)
\, \prod_{p \in {\mathcal P}} (\delta_p , \varphi_p) \, \prod_{p
\notin {\mathcal P}} (\delta_p , \Omega_p)
$$
$$
= \varphi_\infty (0) \, \prod_{p \in {\mathcal P}}  \varphi_p (0)
\, \prod_{p \notin {\mathcal P}}  \Omega_p (0) = \varphi_\infty
(0) \, \prod_{p \in {\mathcal P}}  \varphi_p (0)\, =
\varphi_{{\mathcal P}} (0)  \, .  \eqno(16)
$$

The Gauss integrals, which play an important role in mathematical
physics, may be written in the same form for the real and $p$-adic
case, {\it i.e.}
$$
\int_{{\mathbb Q}_v} \chi_v (a\ x^2 + b\ x) \, dx = \lambda_v (a)
\, |2 a |_v^{-\frac{1}{2}} \, \chi_v\Big( -\frac{b^2}{4 a} \Big)\,
, \quad a \neq 0 \, ,         \eqno(17)
$$
where $v = \infty , \, 2 , \, \cdots , p , \, \cdots$ and
$\lambda_v (a)$ is the definite complex-valued function (see, e.g.
\cite{volovich1}).

\bigskip

\noindent{\bf Proposition.} {\it Gauss integrals satisfy product
formula}
$$
\int_{{\mathbb Q}_\infty} \chi_\infty (a x_\infty^2 + b x_\infty)
\, dx_\infty \, \prod_p \int_{{\mathbb Q}_p} \chi_p (a x_p^2 + b
x_p) \, dx_p   = 1\, , \quad a \in {\mathbb Q}^\ast \, , \, b \in
{\mathbb Q} \, .    \eqno(18)
$$

\noindent{\it Proof. \,}  It follows from the formula (17). A
proof that $\lambda_\infty (a)\, \prod_p \lambda_p (a) = 1$ if $a
\in {\mathbb Q}^\ast$ can be found in Ref 8. It is well known that
adelic multiplicative $|r|^\alpha$ and adelic additive $\chi (r)$
characters are equal to unit if $r \in {\mathbb Q}^\ast$ and $r
\in {\mathbb Q}$, respectively. $\square$

\bigskip

\noindent{\bf Example 6.}   Additive character $\chi (a x^2 + b
x)$, where $a\in {\mathbb A}^\ast$ and $b ,\,  x \in {\mathbb A}$
is an adelic generalized function.

It follows due to the property of integral $\int_{\mathbb A} \chi
(a x^2 + b x) \, \varphi_{\mathcal P} (x)\, dx$ which contains
only finitely many factors different from $ \Omega_p (|b_p|_p)$.

\bigskip
Note that the kernel
$$
K (a, b) = \prod_v \lambda_v (a_v) \, |2 a_v|_v^{-1/2} \, \chi_v
\big( - \frac{b_v^2}{4 a_v} \big) \, , \quad a_v \neq 0, \,\, a ,
b \in {\mathbb A} \, ,   \eqno(19)
$$
is also an adelic generalized function which depends on $b$. Owing
to the formula
$$
\int_{{\mathbb Q}_p} \lambda_p (a_p) \, |2 a_p|_p^{-1/2} \, \chi_p
\big( -\frac{b_p^2}{4 a_p} \big) \,\, \Omega_p (|a_p|_p) \, da_p
$$
$$
= \int_{{\mathbb Z}_p} da_p \, \int_{{\mathbb Q}_p} \chi_p \big(
a_p x^2 + b_p x \big) \, dx = \Omega_p (|b_p|_p)         \eqno(20)
$$
one has the following transform
$$
\Lambda [\varphi_{\mathcal P}] (b) = \int_{\mathbb A} K (a, b)\,
\varphi_{\mathcal P} (a) \, da = \Lambda_\infty [\varphi_\infty]
(b_\infty) \, \prod_{p\in {\mathcal P}} \Lambda_p [\varphi_p]
(b_p) \, \prod_{p\notin {\mathcal P}} \Omega_p (|b_p|_p) \, .
\eqno(21)
$$
The right-hand side of (21) resembles an elementary function on
${\mathbb A}$.

Formula (14) may be regarded as the Mellin transform
$\Phi_{\mathcal P} (\alpha)$ of an elementary function
$\varphi_{\mathcal P} (x)$. $\Phi_{\mathcal P} (\alpha)$ can be
analytically continued on the whole complex plane $\alpha$ except
simple poles in $\alpha = 0$ and $\alpha = 1$. Functional
relation, called the Tate formula, holds:
$$
\Phi_{\mathcal P} (\alpha)= \tilde{\Phi}_{\mathcal P} (1-\alpha)
\, ,     \eqno(22)
$$
where $\tilde{\Phi}_{\mathcal P}$ is the Mellin transform of the
Fourier transform $\tilde{\varphi}_{\mathcal P}$ .

%%%%%%%%

\bigskip
\noindent{\bf 3.\ \  Adelic quantum mechanics}
\bigskip

Since 1987, there has been an intensive and successful activity in
application of $p$-adic numbers in theoretical and mathematical
physics (for a review see, e.g. \cite{volovich1}). One of the
greatest achievements is the formulation of $p$-adic quantum
mechanics \cite{volovich2}. Recently, adelic quantum mechanics has
been introduced \cite{dragovich1} which generalizes and unifies
ordinary and $p$-adic quantum mechanics. It can be defined as a
triple $(L_2 ({\mathbb A}),\, W (z), \, U(t))$, where $L_2
({\mathbb A})$ is the Hilbert space on ${\mathbb A}$, $W (z)$ is a
unitary representation of the Heisenberg-Weyl group on $L_2
({\mathbb A})$, and $U (t)$ is a unitary representation of the
evolution operator on $L_2 ({\mathbb A})$. We shall state here
only some of those aspects of adelic quantum mechanics which are
in close relation to adelic generalized functions. As an
illustration we shall use the harmonic oscillator.

The eigenfunctions may be obtained by solving the equation
$$
\int_{\mathbb A} {\mathcal K}_t (x , y)\, \psi (y)\, dy = \chi (E
t) \, \psi (x) \, , \quad E \,, t \, , x \in {\mathbb A} \, ,
\eqno(23)
$$
where ${\mathcal K}_t (x , y)$ is the kernel of the evolution
operator $U (t)$. In the case of harmonic oscillator one has
$$
{\mathcal K}_t (x , y) = \lambda (2 \, \sin{t}) \,
|\sin{t}|^{-\frac{1}{2}} \, \chi \Big( \frac{x y}{\sin{t}} -
\frac{x^2  +y^2}{2 \, \tan{t}} \Big)  \eqno(24)
$$
$$
{\mathcal K}_0 (x , y) = \delta (x - y) = \delta_\infty (x_\infty
- y_\infty)\, \prod_p \delta_p (x_p - y_p)\, .
$$

An orthonormal set of eigenfunctions for the harmonic oscillator
has the form
$$
\psi (x) = 2^{\frac{1}{4}}\, (2^n\, n!)^{-\frac{1}{2}} \, e^{-\pi
x_\infty^2} \, H_n (x_\infty \, \sqrt{2 \pi}) \, \prod_{p\in
{\mathcal P}} \psi_p (x_p) \, \prod_{p\notin {\mathcal P}}
\Omega_p (|x_p|_p)\, ,                   \eqno(25)
$$
where $H_n \, \, (n= 0 , 1 , \cdots)$ are the Hermite polynomials
and $\psi_p$ are $p$-adic eigenfunctions different from
$\Omega_p$.  It is worth noting that $e^{-\pi x_\infty^2} \, H_n
(x_\infty \, \sqrt{2 \pi}) \in {\mathcal S} ({\mathbb Q}_\infty)
\, , \psi_p \in {\mathcal D} ({\mathbb Q}_p)$ and hence $\psi \in
{\mathcal S} ({\mathbb A})$. In fact, (25) represents some of the
elementary functions (7). Since the Hilbert space $L_2 ({\mathbb
A})$ is complete and infinite dimensional linear space, then the
corresponding Schwartz-Bruhat space ${\mathcal S} ({\mathbb A})$
for the harmonic oscillator is also the complete and infinitely
dimensional one.

The simplest adelic eigenstate is given by
$$
\psi_0 (x) = 2^{\frac{1}{4}}\,  e^{-\pi x_\infty^2} \, \prod_p
\Omega_p (|x_p|_p)                        \eqno(26)
$$
which does not change its form under the Fourier transformation,
{\it i.e.}
$$
\tilde{\psi}_0 (\xi) = 2^{\frac{1}{4}}\,  e^{-\pi \xi_\infty^2} \,
\prod_p \Omega_p (|\xi_p|_p) \, .                        \eqno(27)
$$
According to formula (14) the Mellin transform of (26) is
$$
\Phi (\alpha) = \sqrt{2}\, \Gamma \Big( \frac{\alpha}{2} \Big)\,
\pi^{-\alpha/2} \, \zeta (\alpha)  \, ,    \eqno(28)
$$
where $\Gamma$ is the Euler gamma function. In virtue of the Tate
formula (22) one obtains the well-known functional relation for
the Riemann $\zeta$-function:
$$
\pi^{-\frac{\alpha}{2}} \, \Gamma \Big( \frac{\alpha}{2} \Big) \,
\zeta (\alpha) =  \pi^{\frac{\alpha -1}{2}} \, \Gamma \Big(
\frac{1-\alpha}{2} \Big) \, \zeta (1-\alpha) \, .    \eqno(29)
$$
Formula (29) can be obtained starting from any of eigenfunctions
(25). Thus, by adelic approach we get some connection between
quantum properties of the harmonic oscillator and the zeta
function.

Moreover, the adelic model of harmonic oscillator is an
appropriate physical illustration of analysis over adeles,
including also adelic generalized functions.

Note that formulae (19), (20) and (21) are suitable in the study
of the adelic wave function of the de Sitter model of the universe
\cite{dragovich2}.

At the end, recall that the Dirac $\delta$-function, at first
introduced in ordinary quantum mechanics, induced the appearance
of the theory of generalized functions (distributions). Let adelic
quantum mechanics be a challenge to initiate a systematic
investigation of adelic generalized functions.

%%%%%%%%%%

\end{document}